\documentclass[12pt]{article}
\usepackage{epsfig}
\usepackage{amssymb}

\newcommand{\R}{\mathbb{R}}
\newcommand{\C}{\mathbb{C}}
\newcommand{\be}{\begin{equation}}
\newcommand{\ee}{\end{equation}}
\newcommand{\ben}{\[}
\newcommand{\een}{\]}
\newcommand{\bea}{\begin{eqnarray}}
\newcommand{\eea}{\end{eqnarray}}
\newcommand{\kt}{\rangle}
\newcommand{\br}{\langle}
\newcommand{\ed}{\end{document}}

\newcommand{\pbr}{\prec\!}
\newcommand{\pkt}{\!\succ}

\newcommand{\bi}{\begin{itemize}}
\newcommand{\ei}{\end{itemize}}

\begin{document}

\title{Which operator generates time evolution in Quantum Mechanics?}
\author{\\
Miloslav Znojil
\\
\\
\'{U}stav jadern\'e fyziky AV \v{C}R, 250 68 \v{R}e\v{z}, Czech
Republic\footnote{e-mail: znojil@ujf.cas.cz}}
\date{ }
\maketitle

\begin{abstract}

We attract attention to an interesting family of quantum systems
where the generator $H_{(gen)}$ of time-evolution of wave
functions is {\em not} equal to the Hamiltonian $H$. We describe
the origin of the difference $H_{(gen)}-H$ and interpret it as a
carrier of a compressed information about the other relevant
observables.

\vspace{5mm}

\vspace{2mm}


\end{abstract}


\newpage
 \noindent
In the great majority of the reviews of the history of Quantum
Mechanics (cf., e.g., ref.~\cite{Styer} for illustration) the main
emphasis is usually being laid upon its successful descriptions of
the various {\em not too complicated} systems. The admissible
states $\Phi$ are then often just bound states which lie in the
standard Hilbert space ${\cal H}_{phys}^{(stand)}$ (in what
follows we shall write its elements in the specific format
$|\Phi\pkt$ using the slightly deformed Dirac's kets). Similarly,
the Hamiltonians (denoted by the lower-case $h$) and/or the other
observables are usually chosen as some sufficiently elementary
(say, differential) operators in ${\cal H}_{phys}^{(stand)}$.

In such a setting, one can often make use of the knowledge of the
spectral representation of the Hamiltonian
 \be
 h = \sum_{n=0}^\infty\,|n\pkt\,E_n\,\pbr n|\,
 \label{spec}
 \ee
etc. A breakdown of the idyllic situation is encountered, say, in
the various many-fermion problems where a constructive treatment
of the action of a realistic $h$ on a generic $|\Phi\pkt\,
\in\,{\cal H}_{phys}^{(stand)}$ may prove overcomplicated and, for
various more or less purely technical reason, even prohibitively
difficult. In such a context and in a way known and recommended,
say, in nuclear physics \cite{Geyer}, people usually search for a
suitable mapping of the complicated $h$ on some ``equivalent" but
simpler operator $H$. In general, the latter operator would act in
some other, auxiliary, ``reference" Hilbert space ${\cal
H}^{(ref)}$. Typically \cite{Geyer}, a ``realistic"  multinucleon
Hamiltonian $h$ acting in  ${\cal H}_{phys}^{(stand)}$ can be
assigned an idealized isospectral bosonic partner $H$ acting in
some other Hilbert space ${\cal H}^{(ref)}$.

At present, we may already read about a quickly growing number of
applications of the above idea in several other branches of
quantum theory (cf., e.g., the recent Carl Bender's thorough
review \cite{Carl} of the promising features of the so called
${\cal PT}-$symmetric models in field theory, etc). The use of
several Hilbert spaces clarified also several older puzzles. The
list of successes ranges from the suppression of the negative
Klein-Gordon probabilities \cite{aliKG} up to the ``legalization"
of the spurious states in the Lee model \cite{Lee}. In all of
these settings, a decision of working with the mappings of the
Hamiltonians,
 \be
 h = \Omega\,H\,\Omega^{-1}
 \label{quasihje}
 \ee
relies on the key tacit assumption that the simplicity of the
calculations with the transformed $H$ will more than compensate
for some complications arising due to the presence of the mapping
$\Omega$.

A decisive breakthrough in the appreciation of the merits of the
use of several Hilbert spaces at once came with the Bender's and
Boettcher's discovery \cite{BB} that the reading of the above
eq.~(\ref{quasihje}) can be also inverted. Thus, an innovated
model-building in quantum theory could equally well start from the
choice of some mathematically tractable model $H$ which would be
followed by a suitable transition to its ``intractable" but
``physical" equivalent partner $h$. Of course, all the similar
applications must be based on an appropriate and consistent
quantum-theoretical background. Some of its less usual aspects
will be discussed in what follows.

Firstly, a suitable operator $\Omega$ should be introduced and
understood as a transformer of $h$ into $H$ and {\it vice versa}.
With the left-hand side of eq.~(\ref{quasihje}) considered as
acting on a given ket vector $|\Phi\!\pkt \ \in {\cal
H}^{(stand)}_{phys}$ we have to perceive the modified upper-case
operator $H$ as acting on a ket-vector element $|\Phi\kt$ of {\em
another}, reference space ${\cal H}= {\cal H}^{(ref)}$. Once we
use the standard Dirac's bras and kets we get, up to inessential
multiplication constants,
 \be
 |\Phi\kt := \Omega^{-1}\, |\Phi\pkt\ \in  {\cal H}^{(ref)}\,,
 \ \ \ \ \
 \br\Phi| :=\ \pbr \Phi|\,\left [\Omega^{-1}\right ]^\dagger
 \ \in  \ \left [{\cal
H}^{(ref)} \right ]^\dagger\ \sim\ {\cal H}^{(ref)}\,. \label{3}
 \ee
It is worth noticing that by definition, {\em both } the old and
new spaces are {\em self-dual}. At the same time they are {\em
not} unitary equivalent since, by construction,
 \be
 \pbr \Phi|\Phi'\pkt\ = \br \Phi|\Omega^\dagger\,\Omega
 |\Phi'\kt\,.
 \label{zaved}
 \ee
We see that still another Hilbert space has to be introduced. Let
us denote it by the symbol ${\cal H}_{phys}$ without superscript.
It will {\em share} its ket vectors with the ``intermediate" space
${\cal H}= {\cal H}^{(ref)}$ (where the inner product was $\br
\cdot|\cdot\kt$). In parallel, it will {\em differ} from it by the
innovated definition of its linear functionals,
 \be
 {\cal H}_{phys}^\dagger
 :=\left \{
 \br\!\br \Phi| := \br \Phi| \Omega^\dagger\Omega \,\equiv\,
 \pbr\Phi| \,\Omega
 \ \
 \right \}\,.
 \label{5}
 \ee
In this sense, only the mapping between ${\cal
H}^{(stand)}_{phys}$ and ${\cal H}_{phys}$ can be considered
norm-preserving and unitary.

Once we assume that $H$ acting in ${\cal H}_{phys}$ is our
``simplest" (i.e., e.g., ordinary differential) operator, we may
combine eq.~(\ref{quasihje}) with conventions (\ref{3}) and
(\ref{5}) and write down the spectral representation of $H$,
 \be
 H=\sum_{n=0}^\infty\,
 |\,n
 \kt\,E_n\,\br\!\br\,n
 \,|\,.
 \label{speneher}
 \ee
In an opposite direction, the appropriately normalized
biorthogonal basis used in (\ref{speneher}) can be defined as
composed of the doublets of the left and right eigenvectors of
$H$,
 \ben
 H\,|\,n\kt=E_n\,|\,n\kt\,,\ \ \ \ \
 \br\!\br n|\,H=\br\!\br n|\,E_n\,,\ \ \ \ \ \
 \br\!\br m|\,n\kt=\delta_{mn}\,,\ \ \ \ \ \ \
 I=\sum_{n=0}^\infty\,|\,n\kt\,\br\!\br n|\,.
 \een
Via insertion of eq.~(\ref{spec}) in eq.~(\ref{quasihje}) it is
now easy to deduce that we may have started from the whole family
of the maps $\Omega$ defined by the (presumably, convergent)
infinite series
 \be
 \Omega=\sum_{n=0}^\infty\,|\,n\pkt\,\mu_n\,\br\!\br\,n|\,.
 \label{indent}
 \ee
Here {\em all} the constants $\mu_n\in \C\setminus \{0\}$ are free
parameters. Once we know that $h=h^\dagger$, i.e.,
 \be
 \Omega\,H\,\Omega^{-1}=
 \left [\Omega^{-1}
 \right ]^\dagger\,H^\dagger\,\Omega^\dagger
 \ee
we immediately deduce that
 \be
 H^\dagger=
 \Theta\,H\,\Theta^{-1}\,
 \ee
where we abbreviated
 \be
 \Theta= \Omega^\dagger\,\Omega=\sum_{m=0}^\infty\,
 |m\kt\!\kt\,
 |\mu_m|^2
 \,\br\!\br m\,|=\Theta^\dagger\,.
 \label{identify}
 \ee
The latter formula offers a formal key to the consistent work in
the Hilbert space ${\cal H}_{phys} \,\equiv\,{\cal H}^{(\Theta)}$
since {\em any} mean value of an operator ${\cal O}$ of an
observable must be real,
 \[
 \br a\,|\,\Theta {\cal O}\,|\,a\kt=
 \br a\,|\, {\cal O}^\dagger\,\Theta\,|\,a\kt\,
 \]
so that we have to demand that
 \be
   {\cal O}^\dagger=\Theta\, {\cal O}\,\Theta^{-1}\,.
   \ee
The scope of this general framework of quantum  theory can be
illustrated by the Bessis' toy non-Hermitian Hamiltonian
 \be
 H(g)=p^2+{\rm i}\,g\,x^3\,, \ \ \ \ \ \ \ g > 0\,.
 \label{DB}
 \ee
It possesses the real and discrete spectrum of energies which is
bounded from below \cite{BB,DDT}. Moreover, in spite of its
manifest non-Hermiticity in ${\cal H}^{(ref)}=L_2(\R)$, it has
been shown \cite{cpt} essentially self-adjoint (and, hence,
``fully legal") in the Hilbert space ${\cal H}^{(\Theta)}$ which
differs from $L_2(\R)$ {\em solely} by {\em another definition} of
the inner product between elements $|a\kt$ and $|b\kt$,
 \be
 (a,b)^{(\Theta)}=\br a|\,\Theta\,|\,b\kt\,.
 \label{scalar}
 \ee
Such an introduction of the new space may be treated as a mere
transition to an alternative metric operator
$\Theta=\Theta^\dagger>0$ in the old space.

%

A certain climax of our present letter comes when we address the
problem of the time-evolution in Quantum Mechanics with $\Theta
\neq I$ and with a manifest time-dependence in all our operators.
Once we prepare an initial state as a normalized vector
$|\varphi(t) \pkt \in  {\cal H}_{phys}^{(stand)}$ at $t=0$, we can
rely only on our understanding of the evolution caused by the
auxiliary self-adjoint Hamiltonians $h(t)$.  In particular, we may
immediately solve any time-dependent Schr\"{o}dinger equation
 \be
 {\rm i}\,\partial_t |\varphi(t)\!\pkt \ \ =\ h(t)\,
 |\varphi(t)\!\pkt\,,\ \ \ \ \ \
 |\varphi(t)\!\pkt \ \ = \ u(t)\,|\varphi(0)\!\pkt
 \label{timeq}
 \ee
and we are sure that the related evolution operator given by
equation
 \be
 {\rm i}\partial_t u(t)=h(t)\,u(t)\,
 \label{seh}
 \ee
is {\em certainly} unitary in  ${\cal H}_{phys}^{(stand)}$,
 \ben
 \pbr \varphi(t) \,|\,
 \varphi(t)\pkt=
 \pbr \varphi(0) \,|\,
 \varphi(0)\pkt\,.
 \een
In the next step of our considerations we recollect the pull-backs
$|\Phi(t)\kt=\Omega^{-1}(t)\, |\varphi(t)\!\pkt$ and $\br\!\br
\Phi(t)\,|=\pbr\!\varphi(t)\,|\,\Omega(t)$ carrying, by
assumption, their own time dependence. It is represented,
formally, by the ``right-action" evolution rule
 \be
 |\Phi(t)\kt=U_R(t)\, |\Phi(0)\kt\,,\ \ \ \ \ \
 U_R(t)=\Omega^{-1}(t)\,u(t)\,\Omega(0)
 \ee
accompanied by its ``left-action" parallel
 \be
 |\Phi(t)\kt\!\kt=U_L^\dagger(t)\, |\Phi(0)\kt\!\kt\,,\ \ \ \ \ \
 U_L^\dagger(t)=\Omega^\dagger(t)\,u(t)\,
 \left [\Omega^{-1}(0)\right ]^\dagger\,.
 \ee
The respective non-Hermitian analogues of the Hermitian evolution
rule~(\ref{seh}) are now obtained by the elementary
differentiation and insertions yielding the two separate
differential operator equations
 \be
 {\rm i}\partial_t U_R(t)=
 -\Omega^{-1}(t)
 \left [{\rm i}\partial_t\Omega(t)
 \right ]\, U_R(t)+H(t)\, U_R(t)\,
 \ee
and
 \be
 {\rm i}\partial_t U_L^\dagger(t)=
 H^\dagger(t)\, U_L^\dagger(t)
 +
 \left [{\rm i}\partial_t\Omega^\dagger(t)
 \right ]\,
 \left [
 \Omega^{-1}(t)
 \right ]^\dagger\, U_L^\dagger(t)\,.
 \ee
We are prepared to verify what happens with the norm $ \br\!\br
\Phi(t)\,|\,\Phi(t)\kt $ of states which evolve with time in the
physical space ${\cal H}^{(\Theta)}$. The elementary
differentiation gives
 \bea
 {\rm i}\partial_t\br\!\br \Phi(t)\,|\,\Phi(t)\kt
 ={\rm i}\partial_t
 \br\!\br \Phi(0)\,|\,U_L(t)\,U_R(t)  \,|\,\Phi(0)\kt=
 \\
 =\br\!\br \Phi(0)\,|\,
 \left [ {\rm i}\partial_t
 U_L(t)
 \right ]
 \,U_R(t)  \,|\,\Phi(0)\kt
 +\br\!\br \Phi(0)\,|\,U_L(t)\,
 \left [ {\rm i}\partial_t
 U_R(t)
 \right ]
   \,|\,\Phi(0)\kt=
   \\
 = \br\!\br \Phi(0)\,|\, U_L(t)\,
 \left [ -H(t)+\Omega^{-1}(t)
 \left [{\rm i}\partial_t\Omega(t)
 \right ]
 \right ]
 \,U_R(t)  \,|\,\Phi(0)\kt+\\
 +\br\!\br \Phi(0)\,|\,U_L(t)\,
 \left [ H(t)
  -\Omega^{-1}(t)
 \left [{\rm i}\partial_t\Omega(t)
 \right ]
 \right ]\,U_R(t)
   \,|\,\Phi(0)\kt=0\,.
   \eea
We see that the norm remains constant also in ${\cal
H}^{(\Theta)}$. In the other words, the time-evolution of the
system specified by the quasi-Hermitian and time-dependent
Hamiltonian $H(t)$ is unitary.

Once we abbreviate $\dot{\Omega}(t)\equiv \partial_t\Omega(t)$,
our latter observation can be rephrased as an explicit
specification of {\em the same} time-evolution generator
 \be
 H_{(gen)}(t)=H(t) -{\rm i}\Omega^{-1}(t)
 \dot{\Omega}(t)
 \ee
entering {\em both} the present quasi-Hermitian updates
 \bea
 {\rm i}\partial_t|\Phi(t)\kt
 =H_{(gen)}(t)\,|\Phi(t)\kt\,,
 \label{SEA}\\
 {\rm i}\partial_t|\Phi(t)\kt\!\kt
 =H_{(gen)}^\dagger(t)\,|\Phi(t)\kt\!\kt\,
 \label{SEbe}
 \eea
of the current textbook time-dependent Schr\"{o}dinger equation
for wave functions. Such a confirmation of the preservation of the
overall unitarity of the evolution (at the cost of a not too
difficult and explicit  redefinition $H \to H_{(gen)}$ of the
generator of time evolution which remains the same for both the
left and right action) may be read as really good news for the
theory.

We arrived at the  answer to the question given in the title. In a
concise discussion we should emphasize that such an answer is
quite surprising because the standard connection between the
Hamiltonian and the time evolution of the system is usually
interpreted as a certain ``first principle" of the quantum
dynamics. In such a context our present constructive argument
against the current postulate $H_{(gen)}=H$ should be perceived as
a formal foundation of an apparently counterintuitive innovative
idea that once we admit some dynamically motivated time dependence
of the Hamiltonian itself, we might also contemplate a parallel
assumption of some equally arbitrary, phenomenologically motivated
time dependence of some other observable quantities.

The key reason for the possible mathematical consistency as well
as for a formal acceptability of such a fairly unusual scenario
should be seen in the deep formal ambiguity of the metric
operator. In principle, this observation (made also in refs.
\cite{Geyer} and \cite{ali}) gives us a huge space and freedom for
the {\em time-dependent} variability of the metric
$\Theta=\Theta(t)$. Let us emphasize that in the generic case with
$\dot{H}(t)\neq 0$ the $t-$dependence of $H\neq H^\dagger$ becomes
immediately transferred to its left and right eigenvectors and,
subsequently (i.e., via formula (\ref{identify})), to the metric
$\Theta=\Theta(t)$ itself. Moreover, in parallel to the freedom of
our choice of the $t-$dependence of the Hamiltonian $H(t)$, there
exists, at least in principle, no mathematical obstruction to  an
unrestricted time-dependence freedom introduced directly in all
the complex coefficients $\mu_n=\mu_n(t)$ (cf., once more,
eq.~(\ref{identify})).

We have to admit that we did not really expect that the ``old"
idea of a mapping $\Omega$ between spaces \cite{Geyer} (and, of
course, between Hamiltonian $H$ and its ``suitable" Hermitian
partner $h$) will survive so easily its transfer to the
time-dependent case with $\Omega=\Omega(t)$. At the same time,
having arrived at our final pair of the generalized quantum
time-evolution differential equations (\ref{SEA}) and (\ref{SEbe})
we find them quite natural and consistent. Summarizing our present
results, the Hamiltonian (i.e., energy-operator $H=H(t)$) was
assumed quasi-Hermitian (i.e., $H^\dagger(t) =
\Theta(t)\,H(t)\,\Theta^{-1}(t)$ at some nontrivial metric
$\Theta(t)=\Theta^\dagger(t)>0$). In parallel, some other
quasi-Hermitian operators $A_j$ of observables were assumed to
carry an independent time-dependence (i.e., we were allowed to
demand that $A_j^\dagger(t) = \Theta(t)\,A_j(t)\,\Theta^{-1}(t)$).
Under these assumptions we demonstrated that the evolution of the
system in question still remains unitary.

The latter observation throws new light upon some applications of
the models with $\Theta \neq I$ (cf., e.g., the ones reviewed in
\cite{Carl}) where the values of all the coefficients
$\mu_n=\mu_n(t)$ are fixed using a suitable phenomenological
postulate. In the other applications reviewed in \cite{Geyer} the
variability of $\mu_n$s is also being removed, step by step, via
an appropriate selection of some other operators decided to play
the role of observables. We may conclude that there is nothing
counterintuitive in the manifest ``decoupling" of $H_{(gen)}(t)$
from $H(t)$. It is fairly easily accepted immediately after one
imagines that the information about the quantum evolution {\em
can} be carried {\em not only} by the manifestly time-dependent
Hamiltonian $H(t)$ {\em but also} by some other manifestly
time-dependent observables $A_j(t)$, $j=1,2, \ldots$. Naturally,
the latter information must also influence the overall evolution
of the system.

In each of the eligible scenarios and approaches, the constraints
imposed upon the variability of the coefficients in $\Omega(t)$
may be considered controllable and at our disposal. After an
acceptance of such a concept of the time-dependence in quantum
dynamics, our present main {\em technical} contribution may be
seen in an almost elementary check of the mutual compatibility and
independence of the two ``input-information" hypotheses involving
$H=H(t)$ {\em and} $\Theta=\Theta(t)$.

\section*{Acknowledgement}

Work supported by GA\v{C}R, grant Nr. 202/07/1307, Institutional
Research Plan AV0Z10480505 and by the M\v{S}MT ``Doppler
Institute" project Nr. LC06002.



\ed